\documentclass{elsart}

\usepackage{natbib}

\usepackage{graphicx}
\usepackage{amssymb}

\def\chan{$\it{Chandra}$}
\def\rosat{$\it{ROSAT}$}
\def\asca{$\it{ASCA}$}

\begin{document}

\begin{frontmatter}


\title{A Multi-Wavelength Study of the Western Lobe of W50 Powered by the Galactic Microquasar SS~433}
%\thanks[label1]{}
\author[a]{A. Moldowan},
\author[a]{S. Safi-Harb},
\author[b]{Y. Fuchs},
\author[c]{G. Dubner}
%\ead{moldowan@physics.umanitoba.ca}
\address[a]{Department of Physics \& Astronomy, University of Manitoba, Winnipeg, Manitoba, R3T 2N2, Canada, moldowan@physics.umanitoba.ca, samar@physics.umanitoba.ca}
\address[b]{Service d'Astrophysique, CEA/Saclay, France, yfuchs@discovery.saclay.cea.fr}
\address[c]{IAFE, Institute of Astronomy and Space Physics, Buenos Aires, Argentina, gdubner@iafe.uba.ar}
%\address[e]{Member of the Carrera del Investigador Cient\'\i fico of CONICET, Argentina}

%\title{}


%\author{}

%\address{}

\begin{abstract}
W50 remains the only supernova remnant (SNR) confirmed to harbor a microquasar: the powerful enigmatic source
SS~433. Our past study of this fascinating SNR revealed two X-ray lobes distorting the radio shell as well as
non-thermal X-rays at the site of interaction between the SS~433 eastern jet and the eastern lobe of W50. In this
paper we present the results of a 75~ksec \chan\  ACIS-I observation of the peak of W50-west targeted to 1)
determine the nature of the X-ray emission and 2) correlate the X-ray emission with that in the radio and infrared
domains. We have confirmed that at the site of interaction between the western jet of SS~433 and dense interstellar
gas the X-ray emission is non-thermal in nature. The helical pattern observed in radio is also seen with \chan. No
correlation was found between the infrared and X-ray emission.

\end{abstract}


\end{frontmatter}

% main text
\section{Introduction}
\label{} SS~433 is a peculiar binary system, consisting of a  black hole (as proposed by Lopez et al., 2005) and a massive companion. This system is accreting at a super-Eddington rate, and is expelling two-sided relativistic jets at a velocity of 0.26c. These jets precess in a cone of half-opening angle of 20$^{\circ}$ \citep{M84}.

SS~433 is near the center of W50, a large 2$^{\circ}\times$1$^{\circ}$ nebula stretched in the east-west direction,
and catalogued as an SNR  \citep{G05}.The SS~433/W50 system is the only Galactic object known of its kind, giving
rise to a unique laboratory to study the association between SNRs and black holes as well as the interaction
between relativistic jets and the surrounding medium.

%The elongated morphology of W50 has been attributed to the impact of the SS~433 jets energizing %%%and distorting the shell.
%Because of the ellipsoidal shape of W50, and SS~433 is located near its center,
%it is believed that the jets of SS~433 are interacting with the SNR, causing it to elongate along the jets axis.
%It has also been proposed by \citet{BC1989} that W50 is a bubble blown by
%the jets from SS~433.
%mechanisms behind X-ray binaries, relativistic jets, and supernova remnants.

%SS~433 was first discovered in X-rays by \citet{CGC1975} and independently by \citet{S1976}.
This system has been studied extensively in radio continuum and HI \citep{D1998}, millimetre wavelengths \citep{D2000}, and in X-rays with \rosat\ and \asca\ \citep[][and references therein]{SO1997} and with \emph{RXTE} (Safi-Harb \& Kotani, 2002, Safi-Harb \& Petre, 1999). From this multi-wavelength study, it was concluded that the morphology and energetics of W50 are consistent with the picture of the jets interacting with an inhomogeneous medium and likely hitting a denser cloud in the west.

The \chan\ observation presented here provides the highest resolution X-ray image obtained to date of the bright region of the western lobe of W50. This region was chosen because it coincides with
IR emission and can probe the jet-cloud interaction site. We performed a spatially resolved spectroscopy of this region to primarily determine the nature of the emission and correlate the X-ray emission with  radio and IR observations. The paper is organized as follows. In \S 2, we summarize the \chan\ observation imaging and spectral results and compare them to the \rosat\ and \asca\ data. In \S3, we study the X-ray emission in correlation with the infrared and radio emission, and finally present our conclusions in \S4.

\section{\it{Chandra} \rm Data Reduction and Analysis}

The western lobe of W50 was observed with the ACIS-I chips on board \chan\ on 2003 August 21 at a CCD temperature
of -120$^{\circ}$C. The charge transfer inefficiency was corrected using the APPLY\_CTI tool on the level 1 raw
data. A new level 2 file was then obtained using the standard CIAO 3.0 routines. The final exposure time was
71~ksec.

\subsection{Imaging}

To illustrate the W50 region covered by \chan, we show in Fig.~1 the the radio image of W50 (grey scale), and the
regions covered by observations in infrared (large box) and X-ray (small box). The projection on the sky of the
precession cone axes of the SS~433 jets is also overlayed. The radio image shows that the eastern wing of W50
exhibits a corkscrew pattern, which mimics the precession of the arcseconds-scale jets from SS~433 (Dubner et al.,
1998, Hjellming \& Johnston, 1981). Interestingly, there is a hint of a corkscrew pattern visible in the Chandra
image (Fig. 2 and 3), supporting the conclusion that the SS~433 subarcsecond-scale relativistic jets are affecting
the large scale radio and X-ray emission from W50.

In Fig. 2, we show the energy image in which red corresponds to the soft energy band (0.3-2.4~keV) and blue
corresponds to the hard energy band (2.4-10 keV). In Fig. 3, we show the intensity image in the 0.3-10~keV energy
range. We resolve many point sources in the field (a list of which will be provided elsewhere) and note the knotty
structure of the nebula. The X-ray emission peaks at $\alpha$ (J2000) = 19$^h$ 09$^m$ 42$^s$.86, $\delta$ (J2000) =
05$^{\circ}$ 03$^{\prime}$ 38$^{\prime\prime}$.8.

\begin{figure}[h]
\begin{center}
%\begin{minipage}{3in}
\includegraphics[width=4in]{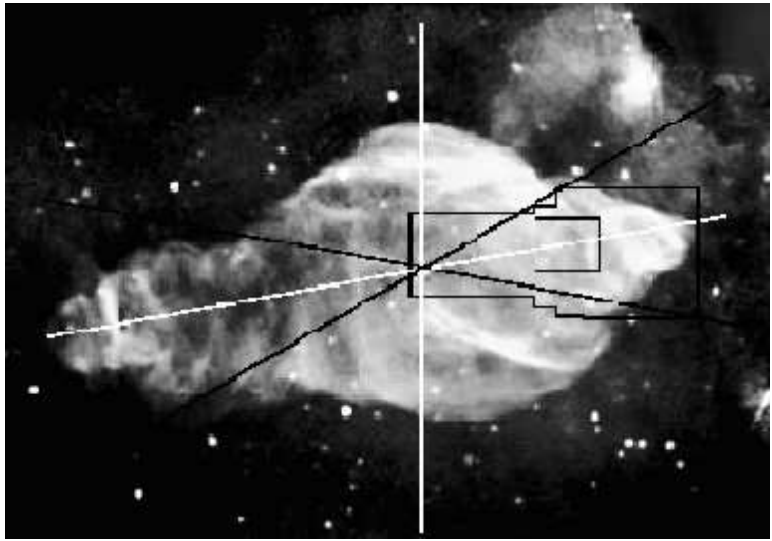}
\caption{The image of W50 in radio. The large and small boxes represent the field of view of the infrared and X-ray observations respectively. The position angle of the jet cone is 100$^{\circ}$ (measured from N to E), and half-opening angle of 20$^{\circ}$}.\\
%\end{minipage}
\end{center}
\end{figure}

\subsection{Spectroscopy}

To perform spatially resolved spectroscopy of the remnant, we excluded the point sources in the \chan\ field, and
extracted spectra from the diffuse emission for 11 regions shown in Fig. 3. The w2 and IRknot2 regions correspond
to the X-ray w2 region presented in \citet{SO1997} and the infrared knot2 region presented by \citet{B1987},
respectively. These regions will be the focus of this paper and are selected in order to compare the \chan\ results
with those found in X-rays with \rosat\ and \asca\ and in infrared with \emph{ISOCAM}.

%\emph{ISOCAM}\footnote {The data are analyzed using ``CIA'', a joint
%development by the ESA Astrophysics Division and the ISOCAM Consortium}.

The proximity of the western lobe to the Galactic plane complicates the spectral analysis because of contamination
by the Galactic ridge. To minimize this contamination, we extracted several background regions from source-free
regions around the diffuse emission from W50 and from the same ACIS chip. We subsequently determined the spectral
parameters using the resulting average background. Spectra were extracted in the 0.5-10.0~keV range. The background
subtracted count rate for the w2 and IRknot2 regions are $\sim(2.3\pm0.05)\times10^{-1}$ counts s$^{-1}$ and
$\sim(6.9\pm0.3)\times10^{-2}$ counts s$^{-1}$ respectively.

To determine whether the emission is thermal or not, we fitted the spectra with thermal bremsstrahlung and
power-law models \citep[following][]{SO1997}. The bremsstrahlung model is characterized by the shock temperature,
$kT$, and the power-law model is characterized by the photon index, $\Gamma$.

\begin{figure} [h]
\begin{center}
%\begin{minipage}
\includegraphics[width=5.5in]{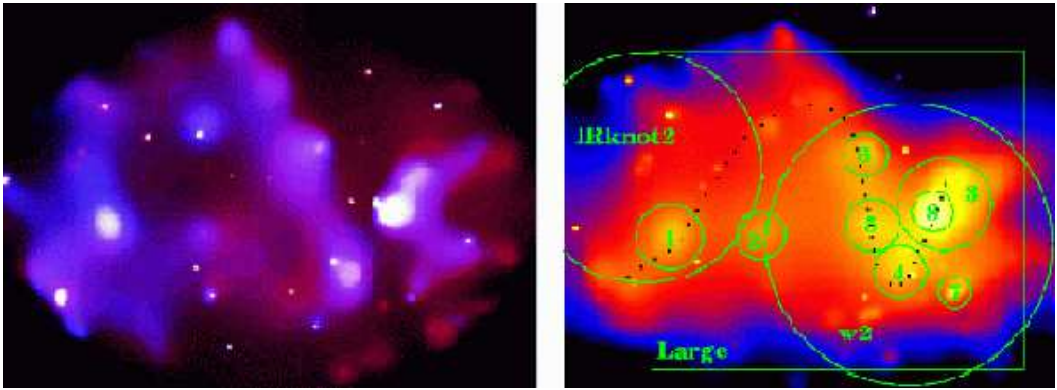}
\bf
\caption{(left): \rm Energy-color image of the western lobe of W50: red=0.3-2.4 keV, blue=2.4-10 keV. Each image was smoothed with a Gaussian with $\sigma$=0$^{\prime\prime}$.5.
\bf Fig. 3. (right): \rm 0.3-10 keV image of W50 showing regions used for spectroscopy (see \S2.2). The dots hint to a corkscrew pattern.}
%\end{minipage}
\end{center}
\end{figure}

\setcounter{figure}{3}
\begin{figure}[h]
\begin{center}
%\begin{minipage}{3in}
\includegraphics[width=4in]{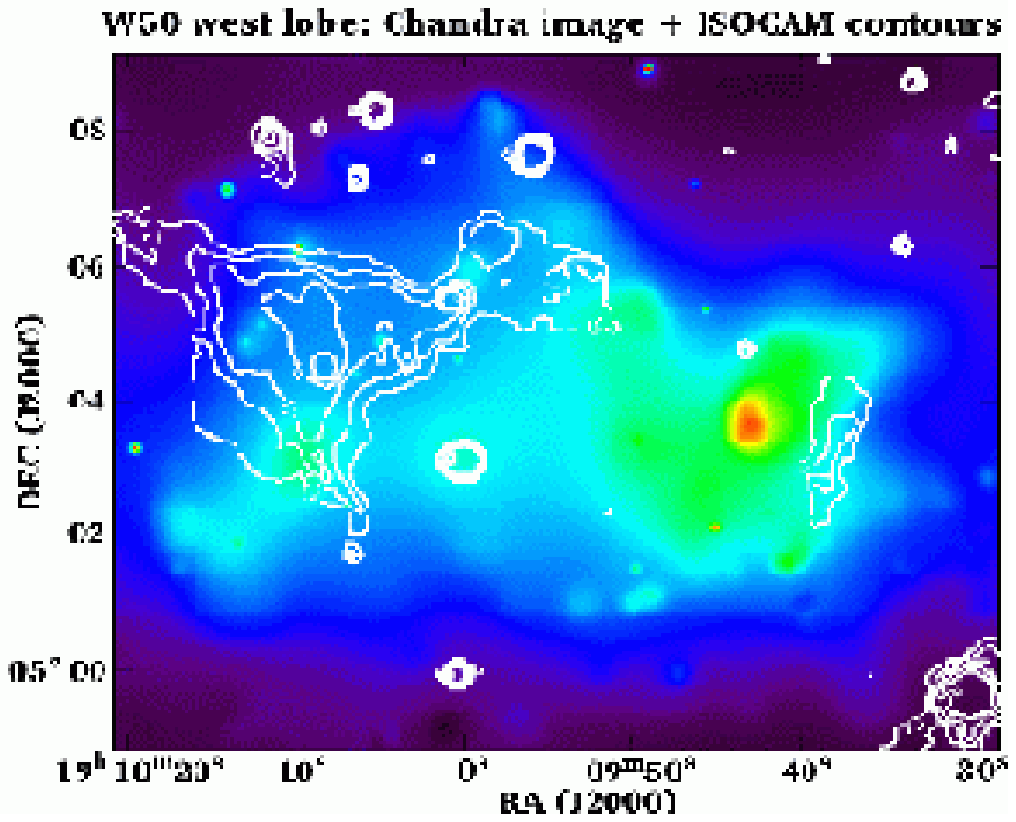}
\caption{X-ray image of the western lobe shown with the infrared contours. See $\S$ 3 for details.}
%\end{minipage}
\end{center}
\end{figure}

Both models give adequate fits in each region. However, we find that the power-law models give slightly lower reduced $\chi^2$ values, and that the temperatures derived from the thermal bremsstrahlung models are high (unrealistically high for even the youngest SNRs). This, together with the absence of line emission in the spectra, leads us to favor the non-thermal interpretation for the X-ray emission. Table~\ref{tab1} summarizes the \chan\ results for the w2 region in comparison to the \rosat\ and \asca\ results. A distance of 3~kpc (scaled by $D_3$) is
used in the luminosity calculations \citep[as in][]{D1998}, and the errors are at the 90\% confidence level.

The spectroscopic results of the other regions are beyond the scope of this paper and will be presented elsewhere;
we note here that the spectrum softens with increasing distance from SS~433  except for regions 3, 6 and 8. This is
consistent with the energy-color image shown in Fig~2.

\begin{table}[h]
\caption{Thermal Bremsstrahlung and Power-law Model Results for the w2 region}
\label{tab1}
\begin{center}
\begin{tabular}{lcccc}\hline\hline
Data Set/ & $N_H$ & $\Gamma$ & $L_x$ & Reduced $\chi^2$ \\
 Model & ($10^{21}$ cm$^{-2}$) & or $kT$ (keV) & ($10^{33}$ $D^2_3$ erg/s) & (DOF) \\\hline
\emph{Chandra} & & & & \\
PL........... & $7.1_{-0.9}^{+1.1}$ & $1.88_{-0.11}^{+0.14}$ & 5.45 & 1.21 (137)\\
BREMS... & $5.8_{-0.7}^{+0.7}$ & $6.07_{-1.13}^{+1.60}$ & 4.56 & 1.21 (137)\\
\emph{ROSAT/ASCA} & & & & \\
PL...........& $5.9_{-1.9}^{+2.3}$ & $2.41_{-0.26}^{+0.34}$ & 5.45 & 1.41 (95)\\
BREMS... & $3.5_{-1.5}^{+1.6}$ & $3.27_{-0.77}^{+1.18}$ & 4.23 & 1.28 (95)\\
\hline
\end{tabular}
\end{center}
\end{table}

We subsequently use the power-law fits to derive the synchrotron emission parameters. Assuming equipartition
between particles and fields and integrating from radio to X-ray frequencies, the equipartition magnetic field
($B_{eq}$), the magnetic energy density ($B_{eq}^2/8\pi$), the total synchrotron electron energy ($U_e$) and the
lifetime of the electrons ($\tau$) can be determined. For the w2 region, we derive
$B_{eq}\sim(2.6-9.8)\times10^{-6}$~G, $B_{eq}^2/8\pi\sim(0.28-3.8)\times10^{-12}$~erg~cm$^{-3}$,
$U_e\sim(0.11-1.6)\times10^{46}$~ergs, and $\tau\sim630-4,515$~years. The range of values corresponds to a=1-100
(the ratio of baryon energy to electron energy). The derived values of the synchrotron parameters as well as $N_H$
(Table~\ref{tab1}) agree with those found using \rosat\ and \asca\ for the w2 region, within error. However, the
spectra appear harder with the \chan\ observation.

%\begin{table}[h]
%\caption{Thermal Bremsstrahlung and Power-law Model Results for the w2 region} \label{tab1}
%\begin{center}
%\begin{tabular}{lcccc}\hline\hline
%Data Set/ & $N_H$ & $kT$ (keV) & $L_x$ & Reduced $\chi^2$ \\
% Model & ($10^{21}$ cm$^{-2}$) & or $\Gamma$ & ($10^{33}$ erg/s $D^2_3$) & (DOF) \\\hline
%\emph{Chandra} & & & & \\
%PL........... & $6.7_{-1.5}^{+1.8}$ & $1.89_{-0.21}^{+0.23}$ & 2.31 & 0.418 (137)\\
%BREMS... & $5.3_{-1.1}^{+1.3}$ & $6.00_{-1.76}^{+2.16}$ & 1.93 & 0.426 (137)\\
%\emph{ROSAT/ASCA} & & & & \\
%PL...........& $5.9_{-1.9}^{+2.3}$ & $2.41_{-0.26}^{+0.34}$ & 5.45 & 1.41 (95)\\
%BREMS... & $3.5_{-1.5}^{+1.6}$ & $3.27_{-0.77}^{+1.18}$ & 4.23 & 1.28 (95)\\\hline
%\end{tabular}
%\end{center}
%\end{table}

\section{Correlation With Radio \& Infrared}

To probe the interaction between the western jet of SS~433 and the ambient medium, we study the X-ray emission in
correlation with radio continuum and HI data obtained with the NRAO\footnote {The NRAO is a facility of the NSF
operated under a cooperative agreement by Associated Universities Inc.} VLA and Green-Bank radio telescope and
infrared data obtained with \emph{ISOCAM}. Fig. 1 shows the radio, infrared, and X-ray regions and Fig. 4 shows the
X-ray emission with the infrared contours.

The average value of $N_H$ found on the basis of the HI observations is $\sim(4-4.4)\times10^{21}$~cm$^{-2}$. This
is slightly lower than the average found using the \chan\ data, which is to be expected.

The energetics in the western lobe found with the X-ray data can then be compared to that found in \citet{D1998}.
We found the total synchrotron electron energy in X-rays to be $\sim2\times10^{45}-3\times10^{46}$~ergs, which is
in good agreement with the energy found from radio observations.

% ($\sim1\times10^{45}-1\times10^{46}$ ergs).

As seen in Fig. 4, there is no correlation between the infrared emission and the peak of X-ray emission. This,
along with the high value of $N_H$ (N$_H$$\geq$2$\times$10$^{22}$~cm$^{-2}$) derived for the IRknot2 region,
suggests that the infrared emission is not associated with the western lobe of W50. The derived value of
$kT\ge4.4$~keV is higher than the expected temperatures for SNRs, indicating that the X-ray emission in IRknot2 is
non-thermal.

\section{Conclusions}

\rm We favor a non-thermal interpretation for the emission of the western lobe of W50. The derived values of $N_H$,
equipartition magnetic field, synchrotron electron energies and lifetimes agree with those derived previously with
\rosat\ and \it{ASCA}\rm.

The infrared emission is not correlated with the peak of X-ray emission. This, in addition to the high value of
$N_H$ derived for this region, suggests that the infrared emission is not originating from W50, and could be
associated with a star forming region (work in progress).

The corkscrew pattern seen in both the radio and X-ray images provides strong support to the hypothesis that the
relativistic jets from SS~433 are causing the morphology of the W50 nebula.

\section{Acknowledgments}
S. Safi-Harb acknowledges support by an NSERC UFA Fellowship and an NSERC Discovery Grant (Canada). Y. Fuchs is
supported by a CNES (France) fellowship. G. Dubner is a Member of the Carrera del Investigador Cient\'\i fico,
CONICET (Argentina).

This research made use of NASA's Astrophysics Data System. We thank the two anonymous referees for their useful
comments.

% \appendix

%\section{}
% \label{}

% Bibliographic references with the natbib package:
% Parenthetical: \citep{Bai92} produces (Bailyn 1992).
% Textual: \citet{Bai95} produces Bailyn et al. (1995).
% An affix and part of a reference:
%   \citep[e.g.][Ch. 2]{Bar76}
%   produces (e.g. Barnes et al. 1976, Ch. 2).

\end{document}